\documentclass{article}

\usepackage{amsmath,amsfonts}

\begin{document}

\title{Group analysis of structure equations for stars in 
radiative and convective equilibrium}

\author{Marek Szyd{\l}owski\\
\small{Astronomical Observatory, Jagiellonian University,} \\
\small{Orla 171, 30-244 Krak\'ow, Poland} 
\and 
Andrzej J. Maciejewski \\
\small{Torun Centre for Astronomy, Nicolaus Copernicus University,}\\ 
\small{Gagarina 11, 87-100 Toru{\'n}, Poland}}

\date{}

\maketitle

\begin{abstract}
It is proposed to use the Lie group theory of symmetries of 
differential equations to investigate the system of equations 
describing a static star in a radiative and convective equilibrium. 
It is shown that the action of an admissible group induces a certain 
algebraic structure in the set of all solutions, which can be used 
to find a family of new solutions. We have demonstrated that, 
in the most general case, the equations admit an infinite parameter 
group of quasi-homologous transformations. We have found invariants 
of the symmetries group which correspond to the fundamental 
relations describing a physical characteristic of the stars such 
as the Hertzsprung-Russell diagram or the mass-luminosity relation. 
In this way we can suggest that group invariants have not only 
purely mathematical sense, but their forms are closely associated 
with the basic empirical relations.
\end{abstract}

\section{Introduction}

A modern theory of stars interiors and their evolution was founded 
on the works of Emden, Lane, Ritter and Kelvin \cite{Chandrasekhar57}. 
They studied equilibrium configurations of polytropic and isothermic gaseous 
spheres. The important results were obtained by Rudzki \cite{Rudzki02} 
who introduced the notion of homologous transformations. Eddington 
\cite{Eddington59} described the effectiveness of radiation transport 
of energy in a star and build the standard model of stellar structure. 
Chandrasekhar \cite{Chandrasekhar57} formulated the homologous theorem 
known as the Stromgren theorem. Schwarzschild \cite{Schwarzschild58} 
described the thermonuclear processes as a source of energy of the stars. 
Some properties of stellar structure can be expressed in terms of 
the Lie group theory of symmetries to obtain the deeper insight in 
structure of solution of equation for stars in radiative and connective 
equilibrium \cite{Biesiada87,Biesiada88}. 

One of the main problems in the group analysis of differential 
equations is the investigation of properties of the group admissible 
by the differential equations structure. In the set of all solutions 
the action of an admissible group induces a certain algebraic 
structure which can be used to find a family of new solutions from 
the known ones 
\cite{Ovsiannikov82,Bluman89,Stefani89,Olver93,Olver95,Barenblatt96}. 

In the present work we apply the Lie group theory to investigate group 
properties of the system of four structure equations describing 
Newtonian static stars in a radiative and convective equilibrium. 
Two of these equations, namely the hydrostatic equilibrium and 
the mass continuity equation, were investigated by Collins from 
the point of view of the group theory \cite{Collins77}. 

As is well known from Stromgren's theorem \cite{Stromgren37}, 
new solutions, for such a system of equations, can be obtained 
from the known ones through the homologous transformations. 
The new solutions describe new configurations with different masses, 
radii, and chemical compositions (the so-called homologous stars). 
We generalize this result by introducing the notion of 
quasi-homologous stars, i.e., stars whose equations of state 
admits quasi-homology symmetries. The homologous stars are a special 
case of the quasi-homologous ones. At the present stage of our 
investigation, this should be treated as a purely mathematical 
result, although it cannot be excluded that the obtained dependencies 
between luminosity and temperature, mass and temperature, and so on 
could be employed in a manner similar to that done by Stromgren 
\cite{Stromgren37} to fully interpret the Hertzsprung-Russell diagram. 

The energy of the star radiated away from its surface is generally 
replenished from reservoirs situated in the very hot central region. 
The transport of energy is possible due to existence of non-vanishing 
temperature gradient in the star. The transfer can occur mainly via 
radiation, conduction and convection. In any case photons, nuclei, 
electrons are exchanged between hotter and cooler parts where 
direction of the energy flow is determined by temperature gradient. 

The problem of application of symmetry group in the context of 
equations coming from astrophysics we considered in our previous 
papers \cite{Biesiada87,Biesiada88,Golda99}. This paper is a continuation 
of new quasi-homologous symmetries which are present in the structural 
equations for stars with convective transport\footnote{Sections 2 and 3 
are the repetition of formulas with necessary corrections from 
Ref.~\cite{Biesiada87} and new formulas (24) and (25) from 
Ref.~\cite{Golda99}, which are crucial to the present analysis}.

\section{Mathematical background}

In the present work we consider differential equation system of 
the following form 
\begin{equation}
\label{eq:1}
\frac{d u^i}{d x} = f^i(x,u^1,\ldots, u^m), \quad i=1,\ldots,m.
\end{equation}
We consider point transformations generated by the infinitesimal 
operator
\begin{equation}
\label{eq:2}
X = \xi(x,u^1,\ldots, u^m) \frac{\partial \phantom{x}}{\partial x} 
+ \sum_{i=1}^{m} \eta^i(x,u^1,\ldots, u^m) 
\frac{\partial \phantom{u^i}}{\partial u^i}.
\end{equation}
For the infinitesimal operator $X$ there is $m$ independent 
invariants which are solutions to the following system
(if we assume analyticity of a field)
\begin{equation}
\label{eq:3}
\frac{d x}{\xi(x,u^1,\ldots, u^m)} = 
\frac{d u^1}{\eta^1(x,u^1,\ldots, u^m)} = \cdots =
\frac{d u^m}{\eta^m(x,u^1,\ldots, u^m)}.
\end{equation}
The point transformation generated by $X$ is called homologous 
if $\xi = ax$ and $\eta^i = g^i u^i$, where $a$, $g^i$ ($i=1,\ldots, m$) 
are constants. Of course, if $X \equiv \lambda^{s}(u) \partial_{s}$ 
then the finite transformation of symmetry $u \to \bar{u}$ are given 
as a solution of equation $d \bar{u}^{s}/d \tau = \lambda^{s}(u)$ 
with the initial conditions $\bar{u}^{s}(\tau = 0)=u^{s}$, $s=1,\ldots,m$. 
It can easily be seen that system (\ref{eq:1}) is similarity invariant 
in this case (for discussion see \cite{Bluman74}). 

A natural extension of this special case leads us to the notion of 
quasi-homologous transformations 
$\xi=\xi(x)$, $\eta^i = \eta^i(u^i)$, $i=1,\dots,m$. 
System (\ref{eq:1}) admits the infinitesimal operator (\ref{eq:2}) 
if and only if the following condition is satisfied 
\begin{equation}
\label{eq:4}
\frac{\partial \eta^i}{\partial x} + \sum_{j=1}^{m} 
\left( \frac{d u^j}{d x} \frac{\partial \eta^i}{\partial u^j} -
\frac{d u^i}{d x}\frac{d u^j}{d x} \frac{\partial \xi}{\partial x} -
\frac{d u^i}{d x} \frac{\partial \xi}{\partial x} \right) - 
X(f^i) = 0.
\end{equation}
Condition (\ref{eq:4}) tells us whether the symmetry operator $X$ 
is admitted by system (\ref{eq:1}) \cite{Ovsiannikov82}. It is easily 
seen that, in the case of a quasi-homologous transformation, equation 
(\ref{eq:4}) assumes form 
\begin{equation}
\label{eq:5}
\frac{d \eta^i}{d u^i} - \frac{d \xi}{d x} = X(\ln f^i).
\end{equation}

Now let us consider the space of independent variable $x$, dependent 
$u^{\alpha}$ and its first derivatives, say $(u^{\alpha})'$ 
($\alpha = 1, \ldots, m$). The action of the Lie group $G$ of the point 
transformation in the space $(x,u)$ can be extended to the space 
$(x,u,u')$. On the other hand an $s$-th order (ordinary or partial) 
differential equation $F(x,u(x),u'(x),\ldots,u^{s}(x))=0$ defines 
a certain manifold $\cal M$ in the space $(x,u,u',\ldots,u^{s})$. 
We say that $F$ is invariant with respect to the group $G$, provided 
that the manifold $\cal M$ is a fixed point with respect to the $s$-th 
extension of $G$, i.e., $G^s({\cal M}) = {\cal M}$. In the terms 
of an infinitesimal operator it means that 
\begin{equation}
\label{eq:6}
X^s F|_{F=0} = 0
\end{equation}
where for the case of first order differential equations 
\begin{align*}
X^s &= X + \xi^{\alpha}_{i}(x,u,u') 
\frac{\partial \phantom{u^{\alpha}}}{\partial u^{\alpha}_{i}} \\
u^{\alpha}_{i} &= \frac{\partial u^{\alpha}}{\partial x^i} \\
\xi^{\alpha}_{i} &= D_{i}(\eta^{\alpha}) - u^{\alpha}_{j} D_{i}(\xi^{j}) \\
D_{i} &= \frac{\partial \phantom{x^i}}{\partial x^i} + 
u^{\alpha}_{i} \frac{\partial \phantom{u^{\alpha}}}{\partial u^{\alpha}}.
\end{align*}

The prolonged operators, which are generators of symmetry of equations 
in the space $(x,u)$, form the structure of the Lie algebra of 
a fundamental group. 

The different method of construction of a partial solution of the system 
without knowledge of fundamental group bases on searching some subgroup 
which is called a similarity group. For which the finite transformations 
are 
\[
\bar{u}^{i} = a^{i} u^{i}, \quad \bar{x}^{\alpha} = a^{n+\alpha} x^{\alpha}, 
\qquad i=1,\ldots,n, \quad \alpha = 1,\ldots,m.
\]
and the Lie algebra is determined by the operator 
\[
X_{i} \equiv u^{i} \frac{\partial \phantom{u^i}}{\partial u^{i}}, \quad 
X_{n+\alpha} = x^{\alpha} 
\frac{\partial \phantom{x^{\alpha}}}{\partial x^{\alpha}}
\]
where all coefficients---dilatation coefficient---$a^{i}, \ldots, a^{n+m}$ 
are positive. It is easy to find such a subgroup because all form 
\[
a^{j} = (a^{1})^{m_{1}j} (a^{2})^{m_{2}j} \ldots (a^{k})^{m_{k}j}, 
\qquad j=k+1, \ldots, N.
\]
The invariants of this subgroup are given 
\[
\Pi = (u^{1})^{\alpha_{1}} (u^{2})^{\alpha_{2}} \ldots 
(u^{N-1})^{\alpha_{N-1}} (u^{N})^{\alpha_{N}}
\]
where $\alpha_{1}, \ldots \alpha_{N}$ are determine from algebraic 
equations. There is a strictly connection between the dimensional 
analysis of equations and describing its fundamental group.

\section{Quasi-homologous transformations of structure equations with 
radiative transport}

The structure for equation for a Newtonian star are as follow 
\cite{Schwarzschild58}
\begin{align}
\label{eq:7}
\frac{d p}{d r} &= - \frac{GM\rho}{r^2} &
\qquad &\textrm{hydrostatic equilibrium} \\ 
\label{eq:8}
\frac{d M}{d r} &= 4\pi r^2 \rho &
\qquad &\textrm{mass continuity} \\ 
\label{eq:9}
\frac{d L}{d r} &= 4 \pi r^2 \epsilon(\rho,T) &
\qquad &\textrm{thermal equilibrium} \\
\intertext{either}
\label{eq:10}
\frac{d T}{d r} &= - \frac{3}{16\pi ac} \frac{\rho}{r^2} 
\frac{L}{T^3} \kappa(\rho,T) & \qquad
&\textrm{radiative equilibrium} \\
\intertext{or} 
\label{eq:11}
\frac{d T}{d r} &= \frac{\Gamma_2 - 1}{\Gamma_2} \frac{T}{p} 
\frac{d p}{d r} = \Gamma \frac{T}{p} \frac{d p}{d r} & \qquad 
&\textrm{adiabatic convective equilibrium}
\end{align}
where $M$ is the mass within the sphere of radius $r$, $\rho$ is the density, 
$p$ is the pressure, $L$ is the luminosity at the surface of the sphere of 
radius $r$, $T$ is the temperature, $\epsilon$ is the energy generation rate, 
$\kappa$ is the opacity, $G$ is the gravitational constant, $c$ is the velocity 
of light, and $a$ is the Stefan-Boltzmann constant. 

First we consider the case of radiative equilibrium. The equation for 
the energy transport through the stellar material can be written as 
a condition for the temperature gradient necessary for the required 
energy flow. It supplies the next basic equation for the stellar structure. 
Assuming the equation of state in the form $p=p(\rho,T)$ we can 
rewrite equation (\ref{eq:7}) in the more convenient form 
\begin{equation}
\label{eq:12}
\frac{d \rho}{d r} = \left( \frac{\partial p}{\partial \rho} \right)^{-1} 
\left(-GM + \frac{3}{16\pi ac} \frac{L}{T^3} \kappa 
\frac{\partial p}{\partial T}\right) \frac{\rho}{r^2}.
\end{equation}
Now, we look for the symmetry transformations of equations 
(\ref{eq:8})--(\ref{eq:10}) and (\ref{eq:12}) generated by the operator 
\begin{equation}
\label{eq:13}
X = \xi(r) \frac{\partial \phantom{r}}{\partial r} + 
\eta^1(\rho) \frac{\partial \phantom{\rho}}{\partial \rho} +
\eta^2(M) \frac{\partial \phantom{M}}{\partial M} + 
\eta^3(L) \frac{\partial \phantom{L}}{\partial L} + 
\eta^4(T) \frac{\partial \phantom{T}}{\partial T}.
\end{equation}

If we denote 
\[
f = -GM + \frac{3}{16 \pi ac} \frac{L}{T^3} \kappa 
\frac{\partial p}{\partial T}
\]
equations (\ref{eq:5}) for quasi-homologous transformations take form 
\begin{align}
\frac{d \eta^{1}}{d \rho} - \frac{d \xi}{r} = 
& - \frac{2\xi}{d r} + \frac{\eta^{1}}{\rho} - 
\eta^{1} \frac{\partial \phantom{\rho}}{\partial \rho} 
\left( \ln \frac{\partial p}{\partial \rho} \right) + 
\frac{\eta^{1}}{f} \left[ \frac{3}{16\pi ac} \frac{L}{T^3} 
\frac{\partial \phantom{\rho}}{\partial \rho} 
\left( \kappa \frac{\partial p}{\partial T}\right) \right] 
- \frac{G \eta^{2}}{f} \nonumber \\
& + \frac{\eta^3}{f} \left( \frac{3}{16\pi ac} \frac{\kappa}{T^3} 
\frac{\partial p}{\partial T}\right)  
+ \frac{\eta^{4}}{f} \left[ -\frac{\partial \phantom{T}}{\partial T} 
\left( \ln \frac{\partial p}{\partial \rho} \right) 
+ \frac{3L}{16\pi ac} 
\frac{\partial \phantom{T}}{\partial T} \left( 
\frac{\kappa}{T^3} \frac{\partial p}{\partial T}\right) \right] 
\label{eq:14} \\
\label{eq:15}
\frac{d \eta^{2}}{d M} - \frac{d \xi}{d r} = 
& \frac{2\xi}{r} + \frac{\eta^{1}}{\rho} \\
\label{eq:16}
\frac{d \eta^{3}}{d L} - \frac{d \xi}{d r} = 
& \frac{2\xi}{r} + \frac{\eta^{1}}{\rho} + \frac{\eta^{1}}{\rho} 
\frac{\partial \epsilon}{\partial \rho} + 
\frac{\eta^{4}}{\epsilon} \frac{\partial \epsilon}{\partial T} \\
\label{eq:17}
\frac{d \eta^{4}}{d T} - \frac{d \xi}{d r} = 
& - \frac{2\xi}{r} + \frac{\eta^{1}}{\rho} + \frac{\eta^{1}}{\kappa} 
\frac{\partial \kappa}{\partial \rho} + \frac{\eta^{3}}{L} + 
\eta^{4} \left( -\frac{3}{T} + \frac{1}{\kappa} 
\frac{\partial \kappa}{\partial T} \right).
\end{align}

Since the right-hand side of equation (\ref{eq:15}) depends only 
on $\rho$, therefore $\eta^{1} = \alpha_{1} \rho$, where $\alpha_{1}$ 
is constant. It is easy to verify, by the same argument and by substitution 
into equations (\ref{eq:14})--(\ref{eq:17}), that also 
\[
\eta^{2} = \alpha_{2} M, \quad 
\eta^{3} = \alpha_{3} L, \quad
\xi = \frac{1}{3}(\alpha_{2} - \alpha_{1})r
\]
where $\alpha_{2}$, $\alpha_{3}$ are constant. By substituting these 
equations into system (\ref{eq:14})--(\ref{eq:17}) one obtains 
\begin{align}
\label{eq:18}
& \left( \frac{d \eta^4}{d T} - \frac{2}{3} \alpha_{2} - 
\frac{4}{3} \alpha_{1} \right) \frac{\partial p}{\partial T} + 
\alpha_{1} \rho \frac{\partial^2 p}{\partial \rho \partial T} + 
\eta^{4} \frac{\partial^2 p}{\partial T^2} = 0 \\
\label{eq:19}
& -\frac{2}{3} \alpha_{2} - \frac{1}{3} + \alpha_{1} \rho -
\frac{\partial \phantom{T}}{\partial \rho} 
\left(\ln \frac{\partial p}{\partial \rho} \right) +
\eta^{4} \ln \frac{\partial p}{\partial T} =0 \\
\label{eq:20}
& \frac{4}{3} \alpha_{1} - \frac{1}{3} \alpha_{2} + \alpha_{3} - 
\frac{d \eta^4}{d T} + \alpha_{1} \frac{\rho}{\kappa} 
\frac{\partial \kappa}{\partial \rho} + \left( -\frac{3}{T} + 
\frac{1}{T} \frac{\partial \kappa}{\partial T} \right) = 0 \\
\label{eq:21}
& (\alpha_{3} - \alpha_{2})\epsilon = \alpha_{1} \rho 
\frac{\partial \epsilon}{\partial T\rho} + \eta^{4} 
\frac{\partial \epsilon}{\partial T}
\end{align}
Equations (\ref{eq:20}) and (\ref{eq:21}) imply that the opacity 
coefficient $\kappa$ and the energy generation rate $\epsilon$ 
are determined by the property of quasi-homologous temperature 
transformation generated by component $\eta^{4}(T)\partial / \partial T$. 
This gives us the following solutions 
\begin{align}
\label{eq:22}
& \phi \left\{ \rho \exp\left( - \alpha_{1} \int_{T_{0}}^{T} 
\frac{d t}{\eta^{4}} \right) , 
\kappa \exp\left[ \int_{T_{0}}^{T} \frac{d t}{\eta^{4}} 
\left( \frac{d \eta^{4}}{d t} - \frac{3\eta^{4}}{t} + 
\frac{\alpha_{2}}{3} - \frac{4\alpha_{1}}{3} - \alpha_{3} 
\right) \right] \right\} = 0 \\
\label{eq:23}
& \psi \left\{ \rho \exp\left( - \alpha_{1} \int_{T_{0}}^{T} 
\frac{d t}{\eta^{4}(t)} \right) , 
\rho \epsilon^{\alpha_{1}/(\alpha_{3} - \alpha_{2})} \right\} = 0
\end{align}
where $\phi$, $\psi$ are arbitrary functions. Instead of form 
(\ref{eq:22}), equivalent and useful forms of representation of 
solutions (\ref{eq:22}) and (\ref{eq:23}) can be used in the 
further analysis \cite{Sneddon57}, namely 
\begin{align}
\label{eq:22a}
\kappa(\rho, T) &= \exp\left[ \int_{T_{0}}^{T} \frac{d t}{\eta^{4}} 
\left( \frac{d \eta^{4}}{d t} - \frac{3\eta^{4}}{t} + 
\frac{\alpha_{2}}{3} - \frac{4\alpha_{1}}{3} - \alpha_{3} 
\right) \right] f\left( \frac{1}{\rho} \exp\left( \alpha_{1} \int_{T_{0}}^{T} 
\frac{d t}{\eta^{4}} \right) \right) \\
\label{eq:23a}
\epsilon(\rho, T) &= \rho^{(\alpha_{3}-\alpha_{2})/\alpha_{1}} 
g\left( \rho^{-(\alpha_{3}-\alpha_{2})/\alpha_{1}}
\exp\left( (\alpha_{3} - \alpha_{2}) \int_{T_{0}}^{T} 
\frac{d t}{\eta^{4}} \right) \right)
\end{align}
where $f$ and $g$ are arbitrary functions. 

It still remains to solve equations (\ref{eq:18}) and (\ref{eq:19}). 
Since we have some freedom in choosing the equation $\eta^{4}(T)$, 
and the function $p(\rho,T)$ is never precisely known, we look for 
those equations of state $p=p(\rho,T)$ that are enforced by the 
transformations generated by operator (\ref{eq:2}). It is easy to 
check by using equations (\ref{eq:18}) and (\ref{eq:19}) that 
$p=p(\rho,T)$ satisfies the continuity condition 
$\partial^{2} p / \partial T \partial \rho$ which in fact is a 
consistency condition for (\ref{eq:18}) and (\ref{eq:19}). 
Depending on the hyperbolic or parabolic character of equations 
(\ref{eq:18}) and (\ref{eq:19}) four cases can be distinguished. 
Let us discuss the solutions for these cases.

\subsection{Case I}
\[
X= \frac{\alpha_{2}-\alpha_{1}}{3} r \frac{\partial \phantom{r}}{\partial r} +
\alpha_{1} \rho \frac{\partial \phantom{\rho}}{\partial \rho} +
\alpha_{2} M \frac{\partial \phantom{M}}{\partial M} + 
\alpha_{3} L \frac{\partial \phantom{L}}{\partial L} + 
\eta^4(T) \frac{\partial \phantom{T}}{\partial T}.
\]
The standard method of reduction to the canonical form, when applied to 
equation (\ref{eq:19}), gives 
\begin{equation}
\label{eq:24}
\frac{\partial^2 p(x,T)}{\partial x \partial T} = 
\frac{4\alpha_{1}/3 + 2\alpha_{2}/3}{\eta^{4}(T)} 
\frac{\partial p}{\partial x}
\end{equation}
where 
\[
x = \rho \exp\left( -  \alpha_{1} \int_{T_{0}}^{T} 
\frac{d t}{\eta^{4}(t)} \right). 
\]
In this case, the general solution of equation (\ref{eq:24}) assumes 
the form \cite{Sneddon57}
\begin{equation}
\label{eq:25}
p(x,T)=h(T) + \int_{x_{0}}^{x} g(\zeta) d \zeta 
\exp \left( \int_{T_{0}}^{T} 
\frac{4\alpha_{1}/3 + 2\alpha_{2}/3}{\eta^{4}(T)} \right)
\end{equation}
where $h(T)$ is of $C^2$ class and $g(\zeta)$ is of $C^1$ class 
of differentiability. 

By substituting the general solution (\ref{eq:25}) in equation 
(\ref{eq:18}) we obtain additional condition for $h(T)$ 
\begin{equation}
\label{eq:26}
\eta^{4}(T) h''(T) + \left( \frac{4}{3}\alpha_{1} + 
\frac{2}{3}\alpha_{2} - \frac{d \eta^{4}}{d T} \right) h'(T) = 0.
\end{equation}
The solution of this equation is 
\begin{equation}
\label{eq:27}
h(T) = C_{1} \int^{T} \exp \left( \int^{\tau} \left( 
\frac{4}{3}\alpha_{1} + \frac{2}{3}\alpha_{2} \right) 
\frac{d t}{d \eta} \right) \frac{d \tau}{\eta^{4}(\tau)} + C_{2}
\end{equation}
where $C_{1}$ and $C_{2}$ are constants. 

This solution implies that in equation (\ref{eq:25}) there is still 
the freedom in choosing a function $g = g(\zeta)$. 

\subsection{Case II}

\[
X= \frac{\alpha_{2}-\alpha_{1}}{3} r \frac{\partial \phantom{r}}{\partial r} +
\alpha_{2} M \frac{\partial \phantom{M}}{\partial M} + 
\alpha_{3} L \frac{\partial \phantom{L}}{\partial L} + 
\eta^4(T) \frac{\partial \phantom{T}}{\partial T}.
\]
By proceeding in the same way as in the previous case we obtain 
\begin{equation}
\label{eq:28}
p(\rho,T) = h(T) + \int^{\rho} g(\rho') \exp \left( \int^{T} 
\frac{2\alpha_{2}}{3\eta^{4}(t)} d t \right) d \rho'
\end{equation}
where
\begin{equation}
\label{eq:29}
h(T) = C_{1} \int^{T} \exp \left( \frac{2}{3} \alpha_{2} 
\int^{\tau} \frac{d t}{\eta^{4}(t)} \right) 
\frac{d \tau}{\eta^{4}(\tau)} + C_{2}
\end{equation}
where $h$, $g$ satisfy differentiability conditions as in the previous 
case and $C_{1}$, $C_{2}$ are constants.

\subsection{Case III}

\[
X= \frac{\alpha_{2}-\alpha_{1}}{3} r \frac{\partial \phantom{r}}{\partial r} +
\alpha_{1} \rho \frac{\partial \phantom{\rho}}{\partial \rho} +
\alpha_{2} M \frac{\partial \phantom{M}}{\partial M} + 
\alpha_{3} L \frac{\partial \phantom{L}}{\partial L}.
\]
Similar calculations give 
\begin{equation}
\label{eq:30}
p(\rho,T) = h(\rho) + \int^{T} \rho^{4/3 + 2\alpha_{2}/3\alpha_{1}} g(t) d t
\end{equation}
where
\begin{equation}
\label{eq:31}
h(T) = C_{1} \rho^{4/3 + 2\alpha_{2}/3\alpha_{1}} + C_{2}
\end{equation}
where $h$, $g$ satisfy the differentiability conditions as previously 
and $C_{1}$, $C_{2}$ are constants. 

\subsection{Case IV}

\[
X= \frac{\alpha_{2}-\alpha_{1}}{3} r \frac{\partial \phantom{r}}{\partial r} +
\alpha_{2} M \frac{\partial \phantom{M}}{\partial M} + 
\alpha_{3} L \frac{\partial \phantom{L}}{\partial L}.
\]
In this case we have the following solutions 
\begin{equation}
\label{eq:32}
p=p(\rho,T), \qquad \alpha_{2}=0.
\end{equation}

We can construct finite transformations and group invariants. 
In case I, for example, we obtain four independent invariants 
$L_{0}$, $M_{0}$, $r_{0}$, and $\rho_{0}$ 
\begin{align}
L_{0} &= L \exp \left( - \alpha_{3} \int^{T} \frac{d t}{\eta^{4}(t)} 
\right), & \qquad & 
M_{0} = M \exp \left( - \alpha_{2} \int^{T} \frac{d t}{\eta^{4}(t)} 
\right) , \nonumber \\
\label{eq:33}
r_{0} &= r \exp \left( \frac{\alpha_{1}-\alpha_{2}}{3} 
\int^{T} \frac{d t}{\eta^{4}(t)} \right), & \qquad & 
\rho_{0} = \rho \exp \left( - \alpha_{1} \int^{T} \frac{d t}{\eta^{4}(t)} 
\right).
\end{align}
By using these invariants, we can construct the new families of 
solutions, for instance 

(i) if $L(T)$ is the solution of (\ref{eq:12}), (\ref{eq:8})
(\ref{eq:9}), and (\ref{eq:10}), then also 
\[
L(E(T)) \exp \left( - \alpha_{3} \int_{E(T_{0})}^{E(T)} 
\frac{d t}{\eta^{4}(t)} \right) 
\]
is the solution, where $E(T)$ is a finite transformation of $T$, 
given from the solution $\bar{T} = E(T)$ of equation 
$d \bar{T} / d \tau = \eta^{4}(\bar{T})$ with the initial condition 
$\bar{T}(\tau =0) = T$.

(ii) if $M(T)$ is the solution of (\ref{eq:12}), (\ref{eq:8})
(\ref{eq:9}), and (\ref{eq:10}), then also 
\[
M(E(T)) \exp \left( - \alpha_{2} \int_{E(T_{0})}^{E(T)} 
\frac{d t}{\eta^{4}(t)} \right) 
\]
is the solution.

\section{Homologous symmetry transformations of structure equations 
with radiative transport}

It is well known that equations (\ref{eq:12}), (\ref{eq:8})
(\ref{eq:9}), and (\ref{eq:10}) admit similarity symmetries for 
the equation of state of ideal gas, $p \sim \rho T$. 
This fact induces a certain class of homologous solutions to 
the system. Some classical results were obtained by Stromgren 
\cite{Stromgren37}. 

In order to investigate how general our results are, let us 
assume the rescaling symmetry $\eta^{4}(T) = \alpha_{4}T$. 
Then, in Case I, for instance, the following equation of state 
is enforced 
\begin{equation}
\label{eq:34}
p(x,T) = \frac{C_{1} T^{(4\alpha_{1} + 2\alpha_{2})/3\alpha_{4}}}%
{\frac{4}{3}\alpha_{1} + \frac{2}{3}\alpha_{2} + \alpha_{4}} 
+ \int_{x_{0}}^{x} g(\zeta) d \zeta 
\frac{T^{1 + (4\alpha_{1} + 2\alpha_{2})/3\alpha_{4}}}%
{\frac{4}{3}\alpha_{1} + \frac{2}{3}\alpha_{2} + \alpha_{4}}.
\end{equation}
One should notice that equation (\ref{eq:34}) contains the following 
form of the equation of state 
\[
p= a \rho^A + b T^B + c \rho^C T^D.
\]

The infinitesimal operator corresponding to the homologous transformations 
takes the form 
\begin{equation}
\label{eq:35}
X= \frac{\alpha_{2}-\alpha_{1}}{3} r \frac{\partial \phantom{r}}{\partial r} +
\alpha_{1} \rho \frac{\partial \phantom{\rho}}{\partial \rho} +
\alpha_{2} M \frac{\partial \phantom{M}}{\partial M} + 
\alpha_{3} L \frac{\partial \phantom{L}}{\partial L} + 
\eta^4(T) \frac{\partial \phantom{T}}{\partial T}.
\end{equation}
The operator given by equation (\ref{eq:35}) has four independent 
invariants, for instance
\[
J_{1}=\rho r^{3\alpha_{1}/(\alpha_{1}-\alpha_{2})}, \quad
J_{2}=\rho M^{-\alpha_{1}/\alpha_{2}}, \quad
J_{3}=L M^{\alpha_{3}/\alpha_{2}}, \quad
J_{4}=L T^{-\alpha_{3}/\alpha_{4}}.
\]
By using these invariants we can arrive at various homology theorems, 
for example the theorem associated with $J_{1}$: if $\rho(r)$ is the 
solution of (\ref{eq:12}) and (\ref{eq:8})--(\ref{eq:10}),
then $\rho\{ r \exp[(\alpha_{2}-\alpha_{1})/3]\}\exp(-\alpha_{1})$ 
is also the solution. 

From $J_{3}$ we obtain $L \sim M^{\alpha_{3}/\alpha_{2}}$ that 
corresponds to the well known Eddington mass-luminosity dependence. 
Infinitesimal operator (\ref{eq:35}) generates a Lie algebra 
spanned by the basic operators 
\begin{equation}
\label{eq:36}
X_{1}= - \frac{1}{3} r \frac{\partial \phantom{r}}{\partial r} +
\rho \frac{\partial \phantom{\rho}}{\partial \rho}, \quad 
X_{2} = \frac{1}{3} r \frac{\partial \phantom{r}}{\partial r} +
M \frac{\partial \phantom{M}}{\partial M}, \quad 
X_{3} = L \frac{\partial \phantom{L}}{\partial L}, \quad
X_{4} = T \frac{\partial \phantom{T}}{\partial T}.
\end{equation}

We now briefly present the basis operators for some particular cases 
that are important from the physical point of view. 

(i) Photon gas, $p \sim T^4$
\begin{equation}
\label{eq:37}
X_{1}= - \frac{r}{3} \frac{\partial \phantom{r}}{\partial r} +
\rho \frac{\partial \phantom{\rho}}{\partial \rho} + 
\frac{T}{3} \frac{\partial \phantom{T}}{\partial T}, \quad 
X_{2} = \frac{r}{3} \frac{\partial \phantom{r}}{\partial r} +
M \frac{\partial \phantom{M}}{\partial M} + 
\frac{T}{6} \frac{\partial \phantom{T}}{\partial T}, \quad 
X_{3} = L \frac{\partial \phantom{L}}{\partial L}, \quad
X_{4} = 0.
\end{equation}

(ii) Ideal gas, $p \sim \rho T$
\begin{equation}
\label{eq:38}
X_{1}= - \frac{r}{3} \frac{\partial \phantom{r}}{\partial r} +
\rho \frac{\partial \phantom{\rho}}{\partial \rho} + 
\frac{T}{3} \frac{\partial \phantom{T}}{\partial T}, \quad 
X_{2} = \frac{r}{3} \frac{\partial \phantom{r}}{\partial r} +
M \frac{\partial \phantom{M}}{\partial M} + 
\frac{T}{6} \frac{\partial \phantom{T}}{\partial T}, \quad 
X_{3} = L \frac{\partial \phantom{L}}{\partial L}, \quad
X_{4} = 0.
\end{equation}

(iii) Degenerate gas, $p \sim \rho^{5/3}$
\begin{equation}
\label{eq:39}
X_{1}= - \frac{r}{6} \frac{\partial \phantom{r}}{\partial r} +
\rho \frac{\partial \phantom{\rho}}{\partial \rho} + 
\frac{M}{2} \frac{\partial \phantom{M}}{\partial M}, \quad 
X_{2} = 0, \quad 
X_{3} = L \frac{\partial \phantom{L}}{\partial L}, \quad
X_{4} = M \frac{\partial \phantom{M}}{\partial M}.
\end{equation}

(iv) Relativistic degenerate electron gas, $p \sim \rho^{4/3}$ 
\begin{equation}
\label{eq:40}
X_{1}= - \frac{r}{3} \frac{\partial \phantom{r}}{\partial r} +
\rho \frac{\partial \phantom{\rho}}{\partial \rho}, \quad 
X_{2} = L \frac{\partial \phantom{L}}{\partial L}, \quad 
X_{3} = 0, \quad
X_{4} = M \frac{\partial \phantom{M}}{\partial M}.
\end{equation}

(v) Ideal and photon gas
\begin{equation}
\label{eq:41}
X_{1}= - \frac{r}{3} \frac{\partial \phantom{r}}{\partial r} +
\rho \frac{\partial \phantom{\rho}}{\partial \rho} + 
\frac{T}{3} \frac{\partial \phantom{T}}{\partial T}, \quad 
X_{2} = 0, \quad 
X_{3} = L \frac{\partial \phantom{L}}{\partial L}, \quad
X_{4} = 0.
\end{equation}

It is interesting to notice that the Eddington mass-luminosity 
dependence is not satisfied any longer, as there are no 
non-trivial invariants associated with $M$. 

(vi) Ideal and degenerate gas 
\begin{equation}
\label{eq:42}
X_{1}= - \frac{r}{6} \frac{\partial \phantom{r}}{\partial r} +
\rho \frac{\partial \phantom{\rho}}{\partial \rho} + 
\frac{M}{2} \frac{\partial \phantom{M}}{\partial M} +
\frac{2T}{3} \frac{\partial \phantom{T}}{\partial T} , \quad 
X_{2} = 0, \quad 
X_{3} = L \frac{\partial \phantom{L}}{\partial L}, \quad
X_{4} = 0.
\end{equation}

\section{Quasi-homologous stars with convective transport}

We start the analysis of stars with convection from equations 
(\ref{eq:7})--(\ref{eq:9}) and (\ref{eq:11}) which describe 
the Newtonian static star with convective transport energy. 
The convective transport of energy means an exchange of energy between 
hotter and cooler layers through the exchange of macroscopic mass
elements, the hotter of which move upwards while the cooler ones 
descend. 

By analogy to the previous analysis instead of equation (\ref{eq:11}) 
it would be useful to operate the new equation that can be 
constructed if we assume the form of the equation of state $p=p(\rho,T)$. 
Then we obtain 
\begin{equation}
\label{eq:43}
\frac{d \rho}{d r} = \left( \frac{\partial p}{\partial \rho} \right)^{-1} 
\left( -1 + \Gamma(\rho,T) \frac{T}{p} \frac{\partial p}{\partial T} 
\right) \frac{GM\rho}{r^2} = 
f\left( \frac{\partial p}{\partial \rho} \right)^{-1} \frac{GM}{r^2}.
\end{equation}
We search for the symmetry operator in form (\ref{eq:13}) for system 
(\ref{eq:7})--(\ref{eq:9}) and (\ref{eq:11}). After extension of operator 
(\ref{eq:13}) on the first derivatives we obtain the equations admissible 
for this operator $X'$ in the form
\begin{align}
\frac{d \eta^{1}(\rho)}{d \rho} - \frac{d \xi(r)}{d r} = 
& X(\ln f^1) = - \frac{2\xi(r)}{r} + \frac{\eta^1(\rho)}{\rho} - 
\eta^4 \frac{\partial \phantom{T}}{\partial T} \ln\left( 
\frac{\partial p}{\partial \rho}\right) + \frac{\eta^2(M)}{M} \nonumber \\ 
& - \eta^1 \frac{\partial \phantom{\rho}}{\partial \rho} \ln\left( 
\frac{\partial p}{\partial \rho}\right) + \frac{\eta^1(\rho)}{f} 
\left( \frac{T}{p} \frac{\partial p}{\partial T} 
\frac{\partial \Gamma}{\partial \rho} + \Gamma \frac{T}{p}
\frac{\partial^2 p}{\partial T \partial \rho} - \frac{\Gamma T}{p^2} 
\frac{\partial p}{\partial T} \frac{\partial p}{\partial \rho} \right) 
\nonumber \\ 
& + \frac{\eta^{4}(T)}{f} \left[ \frac{T}{p} 
\frac{\partial \Gamma}{\partial T} \frac{\partial p}{\partial T} + 
\frac{\Gamma}{p} \frac{\partial p}{\partial T} - \frac{\Gamma T}{p^2} 
\left( \frac{\partial p}{\partial T} \right)^2 + \frac{\Gamma T}{p} 
\frac{\partial^2 p}{\partial T^2} \right] \label{eq:44}\\
\label{eq:45}
\frac{d \eta^{2}(M)}{d M} - \frac{d \xi(r)}{d r} = 
& X(\ln f^2) = \frac{2\xi(r)}{r} + \frac{\eta^1(\rho)}{\rho} \\
\label{eq:46}
\frac{d \eta^{3}(L)}{d L} - \frac{d \xi(r)}{d r} = 
& X(\ln f^3) = \frac{2\xi(r)}{r} + \frac{\eta^1(\rho)}{\rho} + 
\frac{\eta^1}{\epsilon} \frac{\partial \epsilon}{\partial \rho} + 
\frac{\eta^4}{\epsilon} \frac{\partial \epsilon}{\partial T} \\
\frac{d \eta^{4}(T)}{d T} - \frac{d \xi(r)}{d r} = 
& X(\ln f^4) = - \frac{2\xi(r)}{r} + \frac{\eta^2(M)}{M} + \frac{\eta^1}{\rho} + 
\frac{\eta^1}{\Gamma} \frac{\partial \Gamma}{\partial \rho} - 
\frac{\eta^1}{p} \frac{\partial p}{\partial \rho} + 
2\frac{\eta^4}{\Gamma} \frac{\partial \Gamma}{\partial T} \nonumber \\ 
& + \frac{\eta^4}{T} - \frac{\eta^4}{p} 
\frac{\partial p}{\partial T}. \label{eq:47}
\end{align}

From equation (\ref{eq:46}) we obtain $\eta^3(L) = \alpha_{3} L$, where 
$\alpha_3$ is constant and equation (\ref{eq:45}) gives us the linear 
relation $\eta^2(M) = \alpha_{2} M + \beta_{2}$ and $\eta^{1}(\rho) 
= \alpha_{1} \rho$, where $\alpha_1$, $\alpha_2$, $\beta_2$ are 
constant. Then equation (\ref{eq:46}) determines the constraint 
equation in the form 
\begin{equation}
\label{eq:48}
\frac{d \xi(r)}{d r} + \frac{2\xi(r)}{r} = \alpha_{2} - \alpha_{1}
\end{equation}
which can be easily integrated and we obtain 
\begin{equation}
\label{eq:49}
\xi(r) = \frac{\alpha_{2} - \alpha_{1}}{3} r + \frac{C}{r}.
\end{equation}
After substitution of the above solution to equation (\ref{eq:44}) and 
(\ref{eq:47}) we finally obtain that constant $C$ in (\ref{eq:49}) 
must vanish. 

Then from equation (\ref{eq:46}) we obtain 
\begin{equation}
\label{eq:50}
\alpha_{1} \rho \frac{\partial \epsilon}{\partial \rho} + 
\eta^{4}(T) \frac{\partial \epsilon}{\partial T} - 
(\alpha_{3} - \alpha_{2})\epsilon = 0
\end{equation}
which can be integrated by using the standard characteristic method. 
The invariants are determined from the characteristic equation 
\[
\frac{d \rho}{\alpha_{1} \rho} = \frac{d T}{\eta^{4}(T)} = 
\frac{d \epsilon}{(\alpha_{3} - \alpha_{2})\epsilon} 
\]
where we assume that $\alpha_{3} - \alpha_{2}\ne 0$ then 
\begin{equation}
\label{eq:51}
C_{1} = \frac{\rho^{(\alpha_{3} - \alpha_{2})/\alpha_{1}}}{\epsilon}, \qquad 
C_{2} = \frac{\exp\left[(\alpha_{3} - \alpha_{2}) 
\int^{T}\frac{d T}{\eta^{4}(T)}\right]}{\epsilon}.
\end{equation}
It is well known that the general solution of equation (\ref{eq:50}) 
can be given in the form of any function $\psi$ of its invariants, i.e.
\begin{equation}
\label{eq:52}
\psi \left(\frac{\rho^{(\alpha_{3} - \alpha_{2})/\alpha_{1}}}{\epsilon}, 
\frac{\exp\left[(\alpha_{3} - \alpha_{2})\int^{T}
\frac{d T}{\eta^{4}(T)}\right]}{\epsilon} \right) = 0.
\end{equation}
Following Sundman there is the equivalent form of the general solution 
in which the production energy function can be given in the exact form, 
namely 
\begin{equation}
\label{eq:53}
\epsilon = \rho^{(\alpha_{3} - \alpha_{2})/\alpha_{1}}
g\left( \frac{\exp\left[(\alpha_{3} - \alpha_{2})\int^{T}
\frac{d T}{\eta^{4}(T)}\right]}
{\rho^{(\alpha_{3}-\alpha_{2})/\alpha_{1}}} \right).
\end{equation}

From (\ref{eq:44}) or (\ref{eq:47}) we obtain that
\[
\eta^{3}(M) = \alpha_{2} M + \beta_{2}
\]
and $\beta_{2} = 0$. 

Therefore all components of operator $X$ except $\eta^{4}(T)$ are 
determined 
\begin{equation}
\label{eq:54}
\xi(r) = \frac{\alpha_{2} - \alpha_{1}}{3} r, \qquad 
\eta^{1}(\rho) = \alpha_{1} \rho, \qquad 
\eta^{2}(M) =  \alpha_{2} M, \qquad 
\eta^{3}(L) = \alpha_{3} L.
\end{equation}

In our further analysis we assume that that $\Gamma = \Gamma_{0} = 
\mathrm{const}$. This assumption simplifies our calculations but 
we must remember that in general $\Gamma = d \ln T / d \ln p$. 
(If the energy transport is due to radiation and conduction then 
$\Gamma$ must be replaced by $\Gamma_{\mathrm{rad}} = 
3\kappa L p / 16 \pi a c G m T^4$.) This property is destroyed 
if the material function, instead of being products of power 
$p$ and $T$ contains additive terms as in the general case with 
the equation of state. The simplest example is the addition of 
radiation to ideal gas such that $p={\cal R} \rho T/\mu + 
aT^4 / 3 = p_{\mathrm{g}} + p_{\mathrm{r}}$. No strict homology 
(quasi-homology) relation is then possible. But one can try to make 
the enlarged system of equations in which the equation for 
$d \beta / d r$ is added ($\beta = p_{\mathrm{g}}/p$). 
The corresponding equation is 
\begin{equation}
\label{eq:55}
\frac{d \beta}{d r} = \beta \frac{d \phantom{r}}{d r} 
\ln \left( \frac{\cal R}{\mu} \frac{\rho T}{p} \right) = 
\beta \frac{d \phantom{r}}{d r} \ln \left( 
\frac{\rho \Gamma}{p(\rho,T)} \right).
\end{equation}

Now $\beta$ is not constant and in equation (\ref{eq:55}) the relations 
determining $d \rho / d r$, $d T / d r$, and $d p / d r$ should 
be substituted. As a result we obtain that the enlarged system admits 
the homology relation. 

Because we focus on the simplest possible case the problem with the 
constant $\Gamma$ is considered. For $\Gamma = \Gamma_{0}$, equation 
(\ref{eq:44}) assumes the form 
\begin{align}
\frac{\alpha_{2} - \alpha_{1}}{3} =
& \alpha_{2} - \alpha_{1} \rho \frac{\partial \phantom{\rho}}{\partial \rho} 
\ln \left( \frac{\partial p}{\partial \rho} \right) 
- \eta^{4} \frac{\partial \phantom{T}}{\partial T} \left( \ln 
\frac{\partial p}{\partial \rho} \right) 
+ \frac{\alpha_{1}\rho}{f} 
\left( \frac{\Gamma_{0} T}{p} \frac{\partial^2 p}{\partial \rho \partial T} 
- \frac{\Gamma_{0} T}{p^2} \frac{\partial p}{\partial T} 
\frac{\partial p}{\partial \rho} \right) \nonumber \\
\label{eq:56}
& + \frac{\eta^{4}(T)}{f} 
\left[ \frac{\Gamma_{0}}{p} \frac{\partial p}{\partial T} 
- \frac{\Gamma_{0} T}{p^2} \left( \frac{\partial p}{\partial T} \right)^2 
+ \frac{\Gamma_{0} T}{p} \frac{\partial^2 p}{\partial T^2} \right].
\end{align}

Quite similarly we can write equation (\ref{eq:47}) in the constant $\Gamma$ 
and we have 
\begin{equation}
\label{eq:57}
\frac{d \eta^{4}(T)}{d T} - \frac{\alpha_{2} - \alpha_{1}}{3} = 
- \frac{2}{3}(\alpha_{2} - \alpha_{1}) + \alpha_{1} + \alpha_{2} 
+ \frac{\eta^{4}(T)}{T} - \frac{\alpha_{1}\rho}{p} 
\frac{\partial p}{\partial \rho}
\end{equation}
or
\[
\alpha_{1} \rho \frac{\partial p}{\partial \rho} + \eta^{4}(T) 
\frac{\partial p}{\partial T} = \left( - \frac{d \eta^{4}}{d T} 
+ \frac{\eta^{4}}{T} + \frac{4\alpha_{1} + 2\alpha_{2}}{3} 
\right) p.
\]
The solution of equation (\ref{eq:57}) can be given in the form of 
arbitrary function $\psi$ of its invariants, namely 
\begin{equation}
\label{eq:58}
\psi \left( \rho^{-1} \exp \left( \int^{T} \frac{\alpha_{1}}{\eta^{4}} 
d t \right) , 
p^{-1} \exp\left( \int^{T} \frac{d t}{\eta^{4}} \left( 
- \frac{d \eta^{4}}{d t} + \frac{\eta^{4}}{t} + 
\frac{4\alpha_{1} + 2\alpha_{2}}{3} \right) \right) \right) = 0
\end{equation}
or
\begin{equation}
\label{eq:59}
p(\rho, T) = \exp\left( \int^{T} \frac{d t}{\eta^{4}} \left( 
- \frac{d \eta^{4}}{d t} + \frac{\eta^{4}}{t} + 
\frac{4\alpha_{1} + 2\alpha_{2}}{3} \right) \right) 
g \left( \rho^{-1} \exp \left( \int^{T} \frac{\alpha_{1}}{\eta^{4}} 
d t \right) \right).
\end{equation}

After substitution the corresponding terms to (\ref{eq:57}) from 
the exact formula for $p(\rho, T)$, the following condition must 
be fulfilled
\begin{align}
& \frac{d \eta_{4}}{d T} - \frac{\eta_{4}}{T} - \Gamma_{0} T 
\left( -\frac{d \eta_{4}}{d T} + \frac{\eta_{4}}{T} + 
\frac{4\alpha_{1}+2\alpha_{2}}{3} \right) \frac{1}{\eta_{4}}
\frac{d \eta_{4}}{d T} \\
\label{eq:60}
& - \Gamma_{0} T \eta_{4} \frac{\partial \phantom{T}}{\partial T}
\left[ \left( -\frac{d \eta_{4}}{d T} + \frac{\eta_{4}}{T} + 
\frac{4\alpha_{1}+2\alpha_{2}}{3} \right) \frac{1}{\eta_{4}}\right] = 0
\end{align}
The above relation reduces to the simpler form 
\begin{equation}
\label{eq:61}
\frac{d^2 \eta_{4}}{d T^2} + \frac{1-\Gamma_{0}}{\Gamma_{0}} 
\frac{1}{T} \left( \frac{d \eta_{4}}{d T} - \eta_{4} \right) = 0.
\end{equation}
The solution of above equation for any $\Gamma_{0}$ can be obtained 
if we substitute $\eta_{4} \approx T^{x}$. Then 
\begin{equation}
\label{eq:62}
x_{1} = 1, \qquad x_{2} = \frac{\Gamma_{0}}{\Gamma_{0} - 1}
\end{equation}
and finally we 
\begin{equation}
\label{eq:63}
\eta_{4} = C_{1} T + C_{2} T^{(\Gamma_{0}-1)/\Gamma_{0}}
\end{equation}
where $C_{1}$ and $C_{2}$ are constants.

Therefore, we find that for the general form of the equation of state 
the quasi-homology symmetry is admitted by the stellar structure, e.g.\ 
with convective transport, i.e., the operator of this symmetry is 
\begin{equation}
\label{eq:64}
X = \frac{\alpha_{2} - \alpha_{1}}{3} r 
\frac{\partial \phantom{r}}{\partial r} 
+ \alpha_{1} \rho \frac{\partial \phantom{\rho}}{\partial \rho} 
+ \alpha_{2} M \frac{\partial \phantom{M}}{\partial M} 
+ \alpha_{3} L \frac{\partial \phantom{L}}{\partial L} 
+  C_{1} T + C_{2} T^{(\Gamma_{0}-1)/\Gamma_{0}} 
\frac{\partial \phantom{T}}{\partial T}.
\end{equation}
As a special case for $C_{2}=0$ we find that homologous symmetry is obvious 
and $p=T^{(4\alpha_{1}-2\alpha_{2})/3\alpha_{4}} g\left( 
\frac{T^{\alpha_{1}/\alpha_{4}}}{g}\right)$. 

The non-trivial class of quasi-homologous (and homologous) relations 
can be obtained from the invariants. For example, from the invariant 
$J_{1}$ such that  
\begin{equation}
\label{eq:65}
\frac{d L}{\alpha_{3}L} = \frac{d T}{C_{1} T + C_{2} 
T^{(\Gamma_{0}-1)/\Gamma_{0}}}, \qquad
J_{1} = L(T) \exp \left( -\alpha_{3} \int^T 
\frac{d \tau}{C_{1} \tau + C_{2} \tau^{(\Gamma_{0}-1)/\Gamma_{0}}}
\right)
\end{equation}
it is possible to obtain a new class of solutions from the known ones 
which can be treated as a better reconstruction of the main sequence 
on the H-R diagram by using only theoretical assumptions.

\section{Quasi-homologous stars with convective transport}

As the simplest example let us consider star filled by radiation, 
i.e., $p(\rho,T) = aT^4$, $\Gamma = \Gamma_{0} = \frac{1}{4}$, 
$g=1$, then
\begin{equation}
\label{eq:66}
X= \frac{\alpha_{2} - \alpha_{1}}{3} r \frac{\partial \phantom{r}}{\partial r} 
+ \alpha_{1} \rho \frac{\partial \phantom{\rho}}{\partial \rho} 
+ \alpha_{2} M \frac{\partial \phantom{M}}{\partial M} 
+ \alpha_{3} L \frac{\partial \phantom{L}}{\partial L} 
+ (C_{1} T + C_{2} T^{-3}) \frac{\partial \phantom{T}}{\partial T}.
\end{equation}

The important quasi-homologous theorem useful in the generation of 
a solution from the known ones can be formulated in the following way. 
If $L(T)$ is a solution of the stellar structure equation in a convective 
equilibrium, then 
\begin{equation}
\label{eq:67}
L(T') \exp\left( -\alpha_{3} \int^{T'} 
\frac{d \tau}{C_{1} \tau + C_{2} \tau^{-3}} \right) = \mathrm{const}
\end{equation}
is also a solution, where $T' = E(T)$ is a finite transformation which 
is determined as a solution of equation 
\[
\int^{T'} \frac{d T}{C_{1} T + C_{2} T^{-3}} = \tau
\]
and $T'(\tau=0)=T$.

After integration we obtain the exact form of finite transformation 
\[
T' = E(T)\colon C_{1} (T')^{4} + C_{2} = (C_{1} T^{4} + C_{2}) 
e^{4C_{1}\tau}.
\]

This procedure can be repeated in the more general case of constant 
$\Gamma_{0}$. Then we obtain 
\[
T' = E(T)\colon C_{1} (T')^{1/\Gamma_{0}} + C_{2} = (C_{1} T^{1/\Gamma_{0}} 
+ C_{2}) e^{C_{1}\tau/\Gamma_{0}}
\]

\section{Conclusion}

We investigated symmetries for the equations of state of ideal 
gas, polytropic gas, photon gas, degenerated gas, as well as some 
mixed cases, and found the most general form of quasi-homologous 
transformation admitted by the system of equations describing 
Newtonian static stars. In considering the invariants of the 
symmetries group we noticed that they correspond to the relations 
describing physical characteristics of the stars, such as the 
Hertzsprung-Russell diagram or the mass-luminosity relation. 
The group invariants make possible constructing quasi-homologous 
theorems analogous to the Chandrasekhar homologous theorems, 
which provide a prescription for deriving new classes of solutions 
from the present known solutions. 

We characterized, by computing infinitesimal operators, the structure 
of the group admissible by a system of equation describing Newtonian 
static stars in radiative and convective equilibrium. We have shown 
that, in the most general case, the equations admit an infinite 
parameter group of quasi-homologous transformations. These symmetries 
enforce appropriate equations of state. In the particular case of 
a five parameter homologous group of the Stromgren results are 
recovered. 

The equations of state (\ref{eq:34}) and (\ref{eq:59}) and parameterization 
of energy production and capacity are very general. 
They contains both physical and non-physical situations. Our results 
suggest that group invariants have not always purely mathematical 
sense, but that their existence is closely associated with basic 
empirical facts such as the Hertzsprung-Russell diagram or the Eddington 
mass-luminosity dependence. The fact that the main sequence in the 
Hertzsprung-Russell diagram can be reconstructed from the invariants 
proved the effectiveness of the group analysis of fundamental equations 
of astrophysics. We indicated this in two examples on the the main 
sequence models and radiative and convective transport. The main result 
is that except of classical homology there is another type of 
quasi-homologous symmetry which can be applied to certain red giants 
(see \cite[\S 32.2]{Kippenhahn90}). 

Due to existence a new `similarity' between different solutions, the 
quasi-homology relation offer qualitative but the helpful indication for 
interpreting or predicting the numerical solutions.

\section*{Acknowledgments}
The authors thanks dr Biesiada for allowing to use results originally 
going from our collaboration. We are very grateful dr Krawiec for help 
in preparing the paper and stimulating discussion.


\begin{thebibliography}{20}

\bibitem{Chandrasekhar57}
Chandrasekhar S 1957
{\it An Introduction to the Study of Stellar Structure}
(New York: Dover Publ.)

\bibitem{Rudzki02}
Rudzki A 1902
{\it Bull. Astroph.}, {\bf 19} 134

\bibitem{Eddington59}
Eddington A S 1959
{\it The Internal Constitution of the Stars} (New York: Dover Publ.)

\bibitem{Schwarzschild58}
Schwarzschild M 1958
{\it Structure and Evolution of the Stars} 
(Princeton: Princeton University Press)

\bibitem{Biesiada87}
Biesiada M, Golda Z and Szydlowski M 1987
{\it J. Phys. A} {\bf 20} 1313

\bibitem{Biesiada88}
Biesiada M and Szydlowski M 1988
{\it J. Phys. A: Math. Gen.} {\bf 21} 3409

\bibitem{Golda99}
Golda Z and Szydlowski M 1999 
{\it Proceedings of the Eight Marcel Grossman Meeting on General Relativity 
Part A} ed T Piran (Singapore: World Scientific) 330

\bibitem{Ovsiannikov82}
Ovsiannikov L V 1982
{\it Group Analysis of Differential Equations} 
(New York: Academic Press)

\bibitem{Bluman89}
Bluman G W and Kumei S 1989
{\it Symmetries and Differential Equations} 
(New York: Springer-Verlag)

\bibitem{Stefani89}
Stefani H 1989
{\it Differential Equations: Their Solutions Using Symmetries} 
(New York: Cambridge University Press)

\bibitem{Olver93}
Olver P J 1993
{\it Applications of Lie Groups to Differential 
Equations}, 2nd ed. 
(New York: Springer-Verlag)

\bibitem{Olver95}
Olver P J 1995
{\it Equivalence, Invariants and Symmetry} 
(New York: Cambridge University Press)

\bibitem{Barenblatt96}
Barenblatt G I 1996
{\it Scaling, Self-similarity, and Intermediate Asymptotics} 
(New York: Cambridge University Press)

\bibitem{Collins77}
Collins C B 1977
{\it J. Math. Phys.} {\bf 18} 1374

\bibitem{Stromgren37}
Stromgren B 1937
{\it Ery. Exapt. Naturw.} {\bf 16} 465

\bibitem{Bluman74}
Bluman G W and Cole J D 1974
{\it Similarity methods for differential equations} 
(New York: Springer-Verlag)

\bibitem{Sneddon57}
Sneddon I N 1957
{\it Elements of Partial Differential Equations} 
(New York: McGraw-Hill)

\bibitem{Kippenhahn90}
Kippenhahn R and Weigert A 1990
{\it Stellar Structure and Evolution} 
(Berlin: Springer-Verlag)

\end{thebibliography}
\end{document}